\pdfoutput=1
\documentclass[11pt]{arxiv-article}

\bibliography{References}

\usepackage{Definitions}

\usepackage{pifont}
\usepackage{hyperref}
\usepackage{mathrsfs}

\RequirePackage{ytableau}
\ytableausetup{boxsize=.55em,aligntableaux=center}
\newcommand*{\smallday}{\ytableausetup{boxsize=.3em,aligntableaux=center}\ydiagram{1,1}\ytableausetup{boxsize=.55em,aligntableaux=center}}

\makeatletter
\renewcommand*\env@matrix[1][*\c@MaxMatrixCols c]{%
  \hskip -\arraycolsep
  \let\@ifnextchar\new@ifnextchar
  \array{#1}}
\makeatother

\newcommand*{\twobytwo}[4]{\begin{pmatrix}[c|c]
    #1 & #2 \\ \midrule #3 & #4
  \end{pmatrix}}

\newcommand{\OfficialTitle}{
 Charging the Conformal Window
}
\title{\setstretch{1.4}
  {\color{Thoughtless}\Huge\textbf{\dosserif\OfficialTitle}}
}

\hypersetup{pdfauthor={Domenico Orlando and Susanne Reffert and Francesco Sannino},pdftitle={\OfficialTitle},%
  colorlinks=true,linkcolor=ThoughtYouWere,citecolor=ThoughtYouWere,urlcolor=ThoughtYouWere}

\author{%
  \begin{minipage}{.97\linewidth}
    \vspace{1cm}
    \begin{center} \dosserif%
      {\small
         \textbf{Domenico Orlando}\textsuperscript{\ding{71}\ding{72}},
         \textbf{Susanne Reffert}\textsuperscript{\ding{72}} and
         \textbf{Francesco Sannino}\textsuperscript{\ding{73}\ding{74}} 
         }
    \end{center}
    \vspace{1cm}
    \authorBlock{\ding{71}}{\dosserif{} INFN sezione di Torino | Arnold--Regge Center\\
      via Pietro Giuria 1, 10125 Turin, Italy}
    \authorBlock{\ding{72}}{\dosserif{} Albert Einstein Center for Fundamental Physics |
     Institute for Theoretical Physics, University of Bern,\\
     Sidlerstrasse 5, CH-3012 Bern, Switzerland}
   \authorBlock{\ding{73}}{\dosserif{}CP3-Origins \& the Danish Institute for Advanced Study |
     University of Southern Denmark\\ Campusvej 55, DK-5230 Odense, Denmark}
   \authorBlock{\ding{74}}{\dosserif{}Dipartimento di Fisica “E. Pancini”, Università di Napoli Federico II | INFN sezione di Napoli\\ Complesso Universitario di Monte S. Angelo Edificio 6, via Cintia, 80126 Napoli, Italy.}
  \end{minipage}
}

\date{}

\begin{document}

\setstretch{1.2}

\numberwithin{equation}{section}

\begin{titlepage}

  \newgeometry{top=23.1mm,bottom=46.1mm,left=34.6mm,right=34.6mm}

  \maketitle

  \thispagestyle{empty}

  \vfill\dosserif{}

  \abstract{\normalfont{}\noindent{}%
   We investigate the properties of  near-conformal dynamics in a sector of large charge when approaching the lower boundary of the conformal window from the chirally broken phase. To elucidate our approach we use the time-honored example of the phenomenologically relevant \(SU(2)\) color theory featuring $N_f$ Dirac fermions transforming in the fundamental representation of the gauge group. In the chirally broken phase we employ the effective pion Lagrangian featuring also a pseudo-dilaton to capture a possible smooth conformal-to-non-conformal phase transition. We charge the baryon symmetry of the Lagrangian and study its impact on the ground state and spectrum of the theory as well as the would-be conformal dimensions of the lowest large-charge operator. We moreover study the effects of and dependence on the fermion mass term.}

\vfill

\end{titlepage}

\restoregeometry{}

\setstretch{1.2}

\section{Introduction}%
\label{sec:introduction}
The discovery of the Higgs heralds a new era in our understanding of fundamental interactions. It crowns the Standard Model of particle interactions as one of the most successful theories of nature while at the same time it opens new experimental and theoretical possibilities in our quest of a more fundamental theory of nature. It also challenges old and new paradigms. Perhaps one of the most striking features of the Higgs is its near-conformal nature and behavior. 
This is encoded in its couplings and in the lightness of its mass when compared to the one expected to emerge from the infamous quadratic divergences stemming from quantum corrections. 
It would therefore be interesting to see the Higgs emerge as a near-conformal mode in theories free from quadratic divergences. 
One such possibility are gauge-fermion theories with sufficiently large number of fermions near a non-conformal to conformal smooth quantum phase transition. This is a number-of-flavor-driven quantum phase transitions from an \ac{ir} fixed point to a non-conformal phase where chiral symmetry is broken~\cite{Miransky:1996pd}.  

Depending on the underlying mechanism behind the loss of conformality one can envision several scenarios ranging from  a \ac{bkt}-like phase transition discovered in two dimensions~\cite{Kosterlitz:1974sm} and proposed for four dimensions in~\cite{Miransky:1984ef,Miransky:1996pd,Holdom:1988gs,Holdom:1988gr,
Cohen:1988sq,Appelquist:1996dq,Gies:2005as}  to a jumping (non-continuous) phase transition~\cite{Sannino:2012wy}. The subsequent suggestion that theories with a very small number of matter fields in higher-dimensional representations could be (near) conformal~\cite{Sannino:2004qp} culminated in the well-known conformal window phase diagram of~\cite{Dietrich:2006cm} that has served as a road map for lattice studies~\cite{Pica:2017gcb}.   

 In all scenarios, the spectrum is not symmetric on the two sides of the quantum phase transition. In the non-conformal phase, we have a well-defined particle spectrum with states separated by a mass gap. Depending on whether some residual global symmetries are spontaneously broken, the spectrum may feature additional gapless states. In the conformal phase, on the other hand, conformality forbids gaps, enforcing a continuum of  states.  
  
   If the quantum phase transition is smooth, such  as the one due to the annihilation of an \ac{ir} and \ac{uv} fixed point (see~\cite{Sannino:2012wy} for details and a graphical understanding), soon after the transition (annihilation of the fixed points) it is natural to define three regions: a high-energy region dominated by asymptotic freedom,  a quasi-conformal region in which the coupling(s) remain nearly constant, and a low energy one where the theory develops a mass scale.
   Two \ac{rg}-invariant energy scales can be naturally defined: $\Lambda_{UV}$, separating the asymptotically free behavior from the quasi-conformal one, and the scale $\Lambda_{IR}$ below which conformality and, depending on the theory, also certain global symmetries are lost. This behavior is colloquially known as \emph{walking} and it has been invoked several times in the phenomenological literature for models of dynamical electroweak breaking in order to enhance the effect of bilinear fermion operators~\cite{Holdom:1988gs,Holdom:1988gr}.
    The amount of walking is naturally measured in terms of the  \ac{rg} invariant ratio $\Lambda_{UV}/\Lambda_{IR}$. For \ac{qcd}-like theories, this ratio is of order unity while near-conformal theories of \emph{walking} type have ideally $\Lambda_{UV}/\Lambda_{IR} \gg 1$. An equivalent way to view walking is through the emergence of two complex zeros of the beta-function in the near-conformal phase~\cite{Gorbenko:2018ncu}.  

Lattice methods have been developed and proven useful to explore the non-perturbative dynamics of the infrared conformal  window of gauge-fermion theories~\cite{Cacciapaglia:2020kgq}.  The first lattice proof of the existence of a controllably smooth non-perturbative conformal-to-non-conformal quantum phase transition appeared in~\cite{Rantaharju:2019nmh}. 

 Continuous quantum phase transitions may lead to a dilaton-like mode in the broken phase to account for the approximate conformal invariance~\cite{Leung:1985sn,Bardeen:1985sm,Yamawaki:1985zg,Sannino:1999qe,Hong:2004td,Dietrich:2005jn,Appelquist:2010gy}. Its effective action can be implemented following Coleman~\cite{Coleman:1988aos} and it is built by saturating the underlying trace anomaly of the theory associated to the breaking of Weyl invariance. There has been renewed interest in \acp{eft} in these type of actions~\cite{Hong:2004td,Dietrich:2005jn,Goldberger:2008zz,Appelquist:2010gy,Hashimoto:2010nw,Matsuzaki:2013eva,Golterman:2016lsd,Hansen:2016fri,Golterman:2018mfm} in part because lattice studies of \(SU(3)\) gauge theories with $N_f= 8$ fundamental Dirac fermions~\cite{Appelquist:2016viq,Appelquist:2018yqe,Aoki:2014oha,Aoki:2016wnc}, and $N_f = 3$ symmetric 2-index Dirac fermions (sextet)~\cite{Fodor:2012ty,Fodor:2017nlp,Fodor:2019vmw} (known as Minimal Walking Technicolor~\cite{Hong:2004td,Sannino:2004qp,Dietrich:2005jn,Evans:2005pu}) reported evidence of the presence of a light singlet scalar particle in the spectrum. Future studies will reveal if the smooth quantum phase transition observed in~\cite{Rantaharju:2019nmh} also features a dilaton-like state. 
 
 \medskip
In this work we build upon the idea of using the large-charge limit~\cite{Hellerman:2015nra} to gain relevant information about quantum phase transitions~\cite{Orlando:2019skh}. We will make use of  the state-operator correspondence~\cite{Cardy:1984rp,Cardy:1985lth} by assuming it to be approximately valid near the conformal boundary.  

As template, we consider \(SU(2)\) gauge theories with $N_f$ Dirac fermions. These theories are both of major theoretical and phenomenological interest as summarized in a recent review~\cite{Cacciapaglia:2020kgq}. The conformal window is reasonably well understood and confirmed by the agreement among the various lattice simulations. 
In particular, for $N_f\geq 8$, theories are inside the conformal window, while theories with $N_f\leq 4$ break chiral symmetry~\cite{Karavirta:2011zg,Ohki:2010sr,Rantaharju:2014ila,Lewis:2011zb,Hietanen:2013fya,Hietanen:2014xca,Arthur:2016dir, Leino:2018yfd}. The case $N_f=6$ is, however, still not settled~\cite{Karavirta:2011zg,Bursa:2010xn,Hayakawa:2013yfa,Appelquist:2013pqa}. The work of~\cite{Leino:2016njf,Suorsa:2016jsf} for example supports the existence of an~\ac{ir} fixed point at large values of the gauge coupling.  However, for this case larger volumes are required to conclusively decide on the~\ac{ir} fate of the theory.

In particle phenomenology these theories have been extensively used as primary templates for the construction of fundamental theories of dynamical electroweak symmetry breaking. Within the Technicolor paradigm~\cite{Weinberg:1975gm,Susskind:1978ms} they were put forward and investigated in~\cite{Appelquist:1999dq,Duan:2000dy,Ryttov:2008xe}  while within the composite Goldstone Higgs scenario~\cite{Kaplan:1983fs,Kaplan:1983sm}   they were employed in~\cite{Appelquist:1999dq,Cacciapaglia:2014uja}  as well as in~\cite{Katz:2005au,Gripaios:2009pe,Galloway:2010bp,Barnard:2013zea,Ferretti:2013kya}. Additionally, these theories allow for the construction of interesting models of minimal composite dark matter as summarized in~\cite{Cacciapaglia:2020kgq}  in which the resulting dark relic density arises either due to an asymmetry~\cite{Nussinov:1985xr} or via number changing operators~\cite{Carlson:1992fn,deLaix:1995vi,Hochberg:2014dra,Hochberg:2014kqa,Hansen:2015yaa}.  

\medskip
This paper is organized as follows. In Section~\ref{sec:EFT} we introduce the effective action near the lower boundary of the conformal window. The latter features the spectrum of Goldstone Bosons augmented by a dilaton-like state meant to capture the near-conformal dynamics. Here we introduce the fixed baryon charge of the theory and investigate the semi-classical behavior of the theory coupled to a non-trivial gravitational background, going beyond the analysis performed in~\cite{Lenaghan:2001sd}.  Fermion masses are also added and the ground state determined along with its energy.   In Section~\ref{sec:symmetries} we determine the symmetries of the theory by carefully disentangling the ones that break explicitly from the ones that break spontaneously before and after introducing the fixed baryon charge. The spectrum of the theory and the ground state energy are determined in Section~\ref{sec:second-order}. At the fixed point the state-operator correspondence~\cite{Cardy:1984rp,Cardy:1985lth,Hellerman:2015nra}  maps the ground-state energy into the anomalous dimension of the lowest-dimensional operator responsible for the fixed charge.  In our case, working in a near-conformal regime, this gives us the corrections to the putative anomalous dimension as function of the dilaton and fermion mass. These  constitute our main results and can be generalized to any other global symmetry of the theory as well as to other gauge groups such as $SU(3)$. These results offer a new way to explore near-conformal dynamics and its features related, for example, to the non-perturbative dilaton properties as function of the fermion mass operator and anomalous dimension. 
Outlook and conclusions are in Section~\ref{sec:conclusions}.

\section{Effective action and semiclassical description}%
\label{sec:EFT}

Our starting point is the action for a field \(U(t,x)\) transforming in the two-index antisymmetric representation \(\ydiagram{1,1}\) of \(SU(2N_f)\), with a mass term that preserves a \(Sp(2N_f)\) subgroup of \(SU(2N_f)\). The reason for picking this representation will become apparent in Section~\ref{sec:symmetries}.
For ease of notation we will consider only \(N_f\) even; in our conventions the symplectic group is such that \(sp(2) = sl(2)\). The Lagrangian takes the form
\begin{equation}
  L[U] = v^2 \ev*{\del_\mu U \del^\mu U^\dagger } + v^2 m_\pi^2 \ev*{ M U + M^\dagger U^\dagger} - \Lambda_0^4 ,
\end{equation}
where \(v\) accounts for the value of the condensate in the underlying microscopic theory, \(m_\pi \) gives the degenerate masses of the pseudo-Goldstones, and \(\Lambda_0\) is a cosmological constant term that will become important when coupling the theory to a dilaton.

The field \(U(t,x)\) transforms linearly under a chiral rotation as
\begin{align}
  U &\mapsto u U u^t, & u \in SU(2N_f) .
\end{align}
and we chose \(M \) to be the matrix
\begin{equation}
  M = - i \sigma_2 \otimes \Id_{N_f} = \pmqty{ 0 & -1 \\ 1 & 0} \otimes \Id_{N_f} ,
\end{equation}
that preserves the transformations \(u \in Sp(2N_f) \subset SU(2N_f)\) such that \(u^t M u = M\).
This action realizes, at the effective description level, chiral symmetry breaking when the underlying fermions transform according to the fundamental representation of a $Sp(2N)$ group which for $N=1$ is $SU(2)$, i.e. two-color \ac{qcd}. It has been developed and extended over the years~\cite{Appelquist:1999dq,Duan:2000dy,Cacciapaglia:2014uja}.

We are interested in the physics around the lower boundary of the conformal window (for $SU(2)$ gauge group, the bound is around \(N_f \approxeq 6\)).

As has been shown in recent years~\cite{Hellerman:2015nra,Alvarez-Gaume:2016vff,Monin:2016jmo,Loukas:2017lof,Jafferis:2017zna,Favrod:2018xov,Orlando:2019hte,Watanabe:2019adh,Watanabe:2019pdh,Badel:2019oxl,Arias-Tamargo:2019kfr,Alvarez-Gaume:2019biu}, working in sectors of fixed and large charge leads to important simplifications and to an \ac{eft} description of strongly-coupled systems.
It allows us to write an expansion in the inverse of the fixed large charge, giving perturbative control independently of the value of the couplings at cutoff.
The massive theory preserves a \(U(1)\) baryonic symmetry generated by
\begin{equation}
  B = \frac{1}{2} \sigma_3 \otimes \Id_{N_f} = \frac{1}{2} \pmqty{ 1 & 0 \\ 0 & -1}  \otimes \Id_{N_f} ,
\end{equation}
whose corresponding current is
\begin{equation}
  J^B_\mu = 4 i v^2 \ev*{B U \del_\mu U^\dagger} .
\end{equation}
We choose to  describe the behavior of the system in a sector of large charge \(Q = \int\dd{V} J^B_0\).
This introduces a scale \(\Lambda_Q \propto Q^{1/3}\)  consistent with the expectation that in this sector, the transition between the region of broken chiral symmetry and the conformal region can be described as a continuous crossover.

Assuming this behavior, we promote the Lagrangian to be scale-invariant by introducing a dilaton dressing~\cite{Coleman:1988aos}.
Every operator \(\mathcal{O}_k\) of dimension \(k\) is dressed as
\begin{equation}
  \label{eq:dilaton-dressing}
  \mathcal{O}_k \mapsto e^{(k-4) \sigma f} \mathcal{O}_k ,
\end{equation}
where the dilaton \(\sigma\) is a field that transforms non-linearly under scale transformations \(x \mapsto e^\alpha x\) as
\begin{equation}
  \sigma \mapsto \sigma - \frac{\alpha}{f}.
\end{equation}
\(f \) is a dimensionful constant that marks the scale of the breaking of conformal invariance.
In the vicinity of the conformal point we expect the dilaton to have a parametrically small mass \(m_\sigma\).
In the following we want to explore how the masses \(m_\pi\) and \(m_\sigma\) influence the physics near criticality.
Our \ac{eft} is valid in the regime where \(\Lambda_Q\) dominates over any other scales of the problem.

In view of the state/operator correspondence, it is convenient to put the system on a manifold \(\mathcal{M}\) with volume \(V\) and curvature \(R\), so
\begin{equation}
	\Lambda_Q = \left(\frac{Q}{V}\right)^{1/3},
\end{equation}
 and we add the corresponding conformal coupling for \(\sigma\).
All in all, the action is
\begin{multline}
  L[\sigma, U] = v^2 \ev*{\del_\mu U \del^\mu U^\dagger } e^{-2 \sigma f} + v^2 m_\pi^2 \ev*{ M U + M^\dagger U^\dagger}  e^{-y \sigma f} - \Lambda_0^4 e^{- 4 \sigma f} \\
  + \frac{1}{2} \pqty{ \del_\mu \sigma \del_\mu \sigma -  \frac{R}{6 f^2}} e^{-2 \sigma f} - \frac{m_\sigma^2}{16 f^2} \pqty{e^{-4 \sigma f} + 4 \sigma f - 1} + \order{R^2},
\end{multline}
where we have also added terms quadratic in the curvature that do not depend explicitly on the fields.%
\footnote{
If the scales associated to the chiral breaking and conformal breaking are related by \(4N_f v^2 f^2 = 1\), it is possible to rewrite the action in terms of a single field \(V = v U e^{-\sigma f}\):
\begin{multline*}
  L[V] = \ev*{\del_\mu V \del^\mu V^\dagger } - \frac{R}{6 } \ev*{V V^\dagger}  - \frac{\Lambda_0^4}{4 v^4 N_f^2} \ev*{V V^\dagger}^2 
  + \frac{v^{2-y} m_\pi^2}{(2N_f) ^{(y-1)/2}} \ev*{ M V + M^\dagger V^\dagger} \ev*{V V^\dagger}^{(y-1)/2} \\
  - \frac{m_\sigma^2}{2}  \pqty{\frac{\ev*{V^\dagger V}^2}{8 N_f v^2} -  N_f v^2 \log(\frac{\ev*{V^\dagger V}}{2 N_f v^2}) - \frac{N_f v^2}{2}}.
\end{multline*}}
We are not adding the Weyl-anomaly term because we will be mainly interested in the cylindrical geometry, where it vanishes.
In the following, we will neglect the kinetic term \(\| \dd{\sigma}\|^2\) since its only contribution is to add a heavy mode (of order \(\approx \Lambda_Q\)) to the final spectrum~\cite{Orlando:2019skh}.\footnote{One can also integrate out the field \(\sigma\) to get a four-derivative non-linear sigma model description.}

The coefficient \(y\) deserves some extra comments.
We are interested in the effect of the insertion of an operator \(\mathcal{O}_y\) of dimension \(y\) that breaks explicitly the conformal symmetry.
In general, if we add a perturbation of the type \(\delta L = \lambda \mathcal{O}_y\), this will induce a non-derivative interaction for the dilaton of the type~\cite{Rattazzi:2000hs,Goldberger:2008zz}
\begin{equation}
  V_{\mathcal{O}}(\sigma) = e^{- 4 \sigma f } \sum_{n=1}^\infty c_{n}(\mathcal{O}_y) e^{-n (y - 4) \sigma f} ,
\end{equation}
where the coefficients \(c_n\) depend on the details of the perturbation and describe the running of the coupling \(\lambda\).
In the limit of large charge, the higher-order terms are suppressed by inverse powers of \(Q\) and we can keep only the first term in the series.
We are interested in the effect of a mass term that results from the condensation of the fermions in the underlying theory.
It follows that the dimension of the operator is \(y = 3 - \gamma^*\), where \(\gamma^*\) is the anomalous dimension of the condensate.
This is an input parameter in our story and can be estimated for example perturbatively or on the lattice.
The value of \(\gamma^*\) is constrained by causality to be between \(0 < \gamma^* < 1\).
We will discuss the system around the two limit values in detail.

\medskip
For working at large charge, it is convenient to rewrite \(U(t,x)\) as
\begin{equation}
  U(t,x) = e^{i \mu B t} \Sigma(t,x) e^{i \mu B^t t},
\end{equation}
where we have introduced the variable \(\mu\) dual to the baryonic charge.
If we rewrite the action in terms of \(\Sigma(t,x)\) (and use the fact that it is unitary and antisymmetric), we find
\begin{multline}
  \label{eq:mu-Lagrangian}
  L[\sigma, \Sigma] = v^2 \ev*{\del_\mu \Sigma \del^\mu \Sigma^\dagger } e^{-2 \sigma f} +  4 i \mu v^2 \ev*{B \Sigma \del_0 \Sigma^\dagger } e^{-2 \sigma f} \\
  + 2 v^2 \mu^2 \pqty{  \ev*{B \Sigma B^t \Sigma^\dagger} + \ev*{B^2}} e^{-2 \sigma f} + v^2 m_\pi^2 \ev*{ M \Sigma + M^\dagger \Sigma^\dagger } e^{-y \sigma f} \\
  - \Lambda_0^4 e^{- 4 \sigma f} - \frac{ R}{12f^2} e^{-2 \sigma f} - \frac{m_\sigma^2}{16 f^2} \pqty{e^{-4 \sigma f} + 4 \sigma f - 1} + \order{R^2}.
\end{multline}
This action describes the field \(\Sigma\) in a system with a chemical potential (or equivalently, coupled to a background flat connection).
Our independent variable is the baryonic charge \(Q\) which, at the saddle, is obtained as the Legendre transform
\begin{equation}
  Q = \fdv{L}{\mu} .
\end{equation}

Since we are only fixing one charge, we expect the minimum to be homogeneous~\cite{Alvarez-Gaume:2016vff}, so we look for a solution where both \(\sigma\) and  \(\Sigma\) are constant and make the ansatz
\begin{equation}
  \label{eq:ground-state}
  \Sigma(t,x) = \Sigma_0 = E \cos(\varphi) + D \sin(\varphi),
\end{equation}
where
\begin{align}
  E &= i \sigma_2 \otimes \Id_{N_f} , &   D &= \Id_2 \otimes \sigma_2 \otimes \Id_{N_f/2}%
\end{align}
where we are assuming for simplicity that \(N_f\) is even, and \(\varphi\) is an optimization parameter that will be fixed by the \ac{eom}.

The action evaluated at \((\Sigma_0, \sigma_0)\) (where \(\sigma_0\) is the \ac{vev} of the dilaton \(\sigma(t,x) \)) is given by
\begin{multline}
  \label{eq:L0}
  L[\Sigma_0, \sigma_0] = 2 N_f v^2 \mu^2 \sin[2](\varphi) e^{-2 \sigma_0 f} + 4 N_f v^2 m_\pi^2  \cos(\varphi) e^{-y \sigma_0 f} - \Lambda_0^4 e^{-4 \sigma_0 f} \\
  - \frac{ R}{12f^2} e^{-2 \sigma_0 f} - \frac{m_\sigma^2 }{16 f^2} \pqty{ e^{-4 \sigma_0 f} + 4 \sigma_0 f - 1 } + \order{R^2},
\end{multline}
and the corresponding \ac{eom} take the form
\begin{align}
  \fdv{L}{\varphi} =  \fdv{L}{\sigma_0} &=0,  &   \fdv{L}{\mu} &= Q.
\end{align}
Explicitly:
\begin{align}
  &\cos(\varphi) = \frac{m_\pi^2}{\mu^2} e^{(2-y) f \sigma_0}, \\
  &\begin{multlined}
      \frac{ R e^{-2 f \sigma_0 }}{6f} - 4 f m_\pi^2 N_f v^2 y \cos (\varphi ) e^{-f \sigma_0  y} + 4 f  \Lambda^4  e^{-4 f \sigma_0 } - \frac{m_\sigma^2}{4 f} \\
       -4 f \mu^2 N_f v^2 e^{-2 f \sigma_0 } \sin ^2(\varphi )  = 0,
  \end{multlined}\\
  &  4 \mu N_f v^2 e^{-2 f \sigma_0 } \left(1-\frac{m_\pi^4 e^{-2 f \sigma_0  (y-2)}}{\mu^4}\right) = Q,
\end{align}
where \(\Lambda^4 = \Lambda_0^4 + \frac{m_\sigma^2}{16 f^2} \).
The solution for generic values of \(y\) is complicated.
We will instead concentrate on the two extrema of \(y = 3 - \gamma^*\), \emph{i.e.} \(\gamma^* \ll 1\) and \(1 - \gamma^* \ll 1\).

\begin{itemize}
\item For \( \gamma^* \ll 1\), the ground-state energy is
  \begin{multline}
    \label{eq:dimension-small-gamma}
  E %
  = \frac{k_{4/3}}{\tilde V^{1/3}} Q^{4/3} + k_{2/3} \tilde R \tilde V^{1/3} Q^{2/3} + k_0(\mathcal{M})\\
  - \frac{1}{N_f k_{4/3}^4} \pqty{\frac{9m_{\pi}^2}{32v}}^2 \frac{\tilde V^{1/3}}{4\pi^2} \bqty{1 - \gamma^* \pqty{\frac{2}{3}  \log(Q ) - \log(\frac{32 N_f  v^2 \tilde V^{2/3} \pi^2 k_{4/3}}{3} )} } Q^{2/3} \\
  - \Bigg[ \frac{16 \pi^2}{9} N_f k_{2/3} k_{4/3} v^2  m_\sigma^2 \hspace{20em} \\
  - \gamma^* \frac{N_f^2}{3\pi^2 k_{4/3}^5} \pqty{ \frac{9m_{\pi}^2}{32v}}^2 \pqty{\frac{ 5}{8 \pi^2} \frac{N_f^2}{k_{4/3}^4} \pqty{\frac{9m_{\pi}^2}{32v}}^2  -   k_{2/3}  \tilde R } \Bigg] \tilde V   \log(Q) ;
\end{multline}
\item at \((1- \gamma^*) \ll 1\) we find
  \begin{multline}
    \label{eq:dimension-big-gamma}
  E = \frac{k_{4/3}}{\tilde V^{1/3}} Q^{4/3} + k_{2/3} \tilde R \tilde V^{1/3} Q^{2/3} +  k_0(\mathcal{M}) \\
  - \bqty{\frac{16 \pi^2}{9} N_f k_{2/3} k_{4/3} v^2  m_\sigma^2 + ( 1 - \gamma^*) \frac{9 m_\pi^4 }{64 k_{4/3}^3}   }  \tilde V   \log(Q).
\end{multline}
\end{itemize}
Here we have introduced the reduced volume and curvature   \(\tilde V = V/(2\pi^2)\), \(\tilde R = R/6\) and eliminated \(f\) and \(\Lambda\) in favor of the \(k_i\):%
\begin{align}
  k_{4/3} &= \frac{3}{8} \pqty{\frac{\Lambda^2}{\pi  N_f v^2} }^{2/3} , & k_{2/3} &= \frac{1}{4 f^2} \pqty{\frac{\pi^2}{N_f v^2 \Lambda^4} }^{1/3}.
\end{align}
As expected, at the conformal point the spectrum depends only on dimensionless combinations of the original parameters.
The coefficient \( k_0\) depends on the curvature of \(\mathcal{M}\) and, for example, vanishes on the torus \(k_0(T^3) = 0\).%
\footnote{We can decompose \(k_0(\mathcal{M})\) as \(k_0(\mathcal{M}) = k_0^{(1)} R^2 + k_0^{(2)} R_{\mu\nu} R^{\mu\nu} + k_0^{(3)} W^2\), where the \(k_0^{(i)}\) depend on the theory, \(R_{\mu \nu}\) is the Ricci tensor and \(W\) is the Weyl tensor.}

A few remarks are in order.
At the conformal point \(m_\pi = m_\sigma = 0\), the energy depends only on the  dimensionless constants \(k_{4/3}\), \(k_{2/3} \) and \(k_0\) while the non-conformal corrections depend on the background geometry. 
To give them a physical interpretation we can consider the system on a three-sphere of radius \(r_0\), \(\mathcal{M} = S^3(r_0)\): using the state/operator correspondence, the energy of the ground state is mapped to the conformal dimension of the lowest operator of charge \(Q\):
\begin{equation}\label{eq:DeltaQ}
  \Delta(Q) = r_0 E(S^3) =  k_{4/3} Q^{4/3} + k_{2/3} Q^{2/3} + k_0(S^3) = \Delta^*
\end{equation}
In this sense we can think of the \(k_i\) as Wilsonian parameters that encode the details of the theory (they do, for example, depend on \(N_f\)).
It is a very non-trivial result of the large-charge expansion that the spectrum on a generic manifold \(\mathcal{M}\) is directly related to the conformal dimension of the lowest operators.
By construction the \(k_i\) are to be evaluated independently of the \ac{eft}, for example on the lattice (see~\cite{Banerjee:2017fcx,Banerjee:2019jpw} for an analogous study of the \(O(N)\) vector model).
Unsurprisingly, at this point all the dimensionful parameters are subsumed into dimensionless quantities (since we are describing a \ac{cft} that cannot have intrinsic scales).
Note that the result~\eqref{eq:DeltaQ} is purely classical as no quantum corrections to the energy of the classical ground state have been computed.
Their first contribution will appear at  \textsc{nnlo} where it can be subsumed in $k_0$.

Away from the conformal point, the typical signature of the breaking of conformal invariance is the presence of terms that scale logarithmically with the charge~\cite{Orlando:2019skh}.
The couplings of the parameters \(m_\sigma\), \(m_\pi\) and \(y\) have different scalings with respect to charge, volume and number of flavors.
This can be used effectively to identify and separate the single contributions in an independent study of the model (\emph{e.g.} perturbative or on the lattice).

\section{Symmetries}%
\label{sec:symmetries}

Although we have in mind two-color \ac{qcd} with \(N_f\) flavors in the confining phase, the pattern of chiral symmetry breaking that we will outline below is valid for any gauge-fermion theory in which the fermions transform in a pseudoreal representation of the underlying gauge group~\cite{Peskin:1980gc,Preskill:1980mz}. 
The expected \(SU(N_f) \times SU(N_f) \times U(1)\) symmetry is enhanced to \(SU(2N_f) \times U(1)_A\), where the \(U(1)_A\) is anomalous.
In the standard scenario, a quark-quark condensate forms and breaks the symmetry spontaneously to the maximal diagonal subgroup \(Sp(2N_f)\).

The condensate describes a pair of quarks in the microscopic theory, so it must have the form
\begin{equation}
  \label{eq:initial-condensate}
  U^{ff' } = Q_\alpha^{c f} Q_{\alpha'}^{c'f'} \epsilon^{\alpha\alpha'} \epsilon_{cc'}
\end{equation}
where \(\alpha\), \(\alpha'\) are Lorentz (spin) indices, \(c\), \(c'\) are color indices and \(f \), \(f'\) are flavor indices.
It follows that \(U\) transforms in the representation \(\ydiagram{1,1}\) that we have used in the previous section.
The condensate breaks the symmetry spontaneously:
\begin{equation}
  SU(2N_f) \leadsto Sp(2N_f)
\end{equation}
(here and in the following we use the notation \(\leadsto\) to indicate a spontaneous breaking of the symmetry). By Goldstone's theorem, the low-energy spectrum is described by
\begin{equation}
  \dim\left(\frac{SU(2N_f)}{Sp(2N_f)}\right) = 2 N_f^2 - N_f - 1 
\end{equation}
Goldstone \acp{dof} that sit in the representation \((0,1,0,\dots,0)\) of the unbroken \(Sp(2N_f)\), which we will indicate again with a slight abuse of notation as \(\ydiagram{1,1}\).

We are interested in the behavior of the theory at the boundary of the conformal window.
Since we generically expect the system to be strongly coupled, we use the large-charge approach and study it in sectors of fixed baryonic charge.
As discussed in the previous section, this gives an effective chemical potential \(\mu\) (see the Lagrangian in Eq.~\eqref{eq:mu-Lagrangian}) that:
\begin{enumerate}
\item breaks explicitly the global symmetry as 
\begin{equation}
	SU(2N_f) \to S(U(N_f) \times U(N_f)) = SU(N_f)_L \times SU(N_f)_R \rtimes U(1)_B,
\end{equation}
since it only preserves transformations of \(\Sigma\) that commute with \(B\),
  \begin{equation}
    \ev*{B \Sigma \del_0 \Sigma^\dagger} = \ev*{B (u \Sigma u^t) \del_0(u \Sigma u^t)^\dagger} = \ev*{u^\dagger B u \Sigma \del_0 \Sigma^\dagger} \Rightarrow \comm{u}{B} = 0 ,
  \end{equation}
  and have the form
  \begin{align}
    \label{eq:preserved-by-mu}
   u &= \twobytwo{A}{0}{0}{B}, & A, B &\in U(N_f), &\det(A)  \det(B) &= 1 ;
  \end{align}
\item breaks explicitly the conformal symmetry to the rotation group times time translation.
\end{enumerate}
There is an equivalent description of this effect in terms of massive pseudo-Goldstone bosons that non-linearly realize the full symmetry.
In particular, the dilaton in Eq.~\eqref{eq:dilaton-dressing} is the pseudo-Goldstone field that realizes the conformal symmetry.
Since all these fields have a mass of order \(\mu\) fixed by the group structure, they do not contribute to the low-energy spectrum.

Moreover, we are interested in the effect of a mass term that preserves the \(Sp(2N_f)\) symmetry and gives a mass \(m_\pi\) to all the pions in the confining phase.
The mass term  breaks explicitly the symmetry preserved by \(\mu\) as \(S(U(N_f) \times U(N_f)) \xrightarrow[m]{} U(N_f)\).
To see this, we need to impose the invariance of the coupling \(\ev*{MU}\) in the action.
In terms of the matrices \(A\) and \(B\) in Eq.~\eqref{eq:preserved-by-mu} we have
\begin{align}
  \ev*{MU} &= \ev*{M u U u^t}, & \text{where } M &= \pmqty{0 & -1 \\ 1 & 0} \otimes \Id_{N_f}.
\end{align}
Using the cyclicity of the trace this implies
\begin{equation}
  A^t B = 1,
\end{equation}
and since \(A\) and \(B \) are both unitary, we find that the mass term preserves transformations of the type
\begin{equation}
  u = \twobytwo{A}{0}{0}{A^*}
  \in U(N_f),
\end{equation}
which transform in the representation \(\ydiagram{1} \oplus \overline{\ydiagram{1}}\) of \(U(N_f)\).
(Note that the baryonic \(U(1)\) operator \(e^{i \mu B}\) has precisely this form).
It is also convenient to write \(u\) in terms of the algebra as \(u =e^{iv}\) and the corresponding pushforward of the representation is
\begin{align}
  v &= \twobytwo{\pi}{0}{0}{-\pi^t}, & \pi \in u(N_f) .
\end{align}
The action in Eq.~\eqref{eq:mu-Lagrangian} then describes a system with the following explicit symmetry breaking:
\begin{equation}
  SU(2N_f) \xrightarrow[\mu]{}  S(U(N_f) \times U(N_f))  \xrightarrow[m]{} U(N_f) .
\end{equation}
Now we need to describe the effect of the spontaneous breaking for the system   around the ground state in Eq.~\eqref{eq:ground-state}.
The physical expectation is that a condensate will still develop, but this time the standard condensate in Eq.~\eqref{eq:preserved-by-mu} will be broken by the chemical potential into a left and a right diquark \(U^{ff'}_L\)  and \(U^{ff'}_R\), so that each of the \(SU(N_f)\) factors in \(SU(N_f)_L \times SU(N_f)_R \rtimes U(1)_B\) is broken spontaneously to \(Sp(N_f)\).
The condensate will also break the \(U(1)_B\) symmetry, since it has a non-zero baryonic charge.\footnote{More precisely, the condensate breaks spontaneously the product of \(U(1)_B\) and time translations to a linear combination of the two symmetries~\cite{Monin:2016jmo}.}

For \(m_\pi \neq 0\), only a linear combination of the left and right diquarks survives, so that the global \(U(N_f)\) symmetry is broken spontaneously to \(Sp(N_f)\).
All in all the symmetry breaking pattern is
\begin{equation}
  SU(2N_f) \xrightarrow[\mu]{}  S(U(N_f) \times U(N_f))  \xrightarrow[m]{} U(N_f) \leadsto Sp(N_f) ,
\end{equation}
and as a result we expect
\begin{equation}
  \dim\left(\frac{U(N_f)}{Sp(N_f)} \right) = \frac{N_f^2 - N_f}{2}
\end{equation}
Goldstone \ac{dof} that transform in the representation \(\mathbf{1} \oplus \ydiagram{1,1}\) of the unbroken \(Sp(N_f)\).
In terms of \(\pi\), the broken generators are the baryonic \(u(1)_B\) and the generators that satisfy \(\pi \mathcal{I} - \mathcal{I} \pi^t = 0\). 

Fixing the charge breaks Lorentz invariance, which means that the spectrum of the Goldstone fields is richer than in a standard Lorentz-invariant theory.
In general, we expect the presence of type-I Goldstones with linear dispersion relation and of type-II Goldstones with quadratic dispersion.
However, in the case at hand we can exclude the presence of the latter with a simple counting argument.
Each type-II Goldstone field is made of two canonically conjugate degrees of freedom.\footnote{More geometrically speaking, this requires the coset to have a pre-symplectic structure~\cite{Watanabe:2012hr}. %
  We will show explicitly in the next section that this is not the case for this model.}
This means that in order to have type-II Goldstones, the dimension of the coset (which counts the number of \ac{dof}) must be strictly greater than the number of fields.
In our case, we have  \(\dim(\frac{U(N_f)}{Sp(N_f)} )= (N_f^2 - N_f)/2 \) \ac{dof}. On the other hand, the dimension of the representation in which the pions transform is precisely \(\dim(\mathbf{1} \oplus \ydiagram{1,1}) = 1 + (N_f + 1 )(N_f -2)/2\).
There is precisely one field per \ac{dof}, so each field must be of type I and have a linear dispersion relation.
At the conformal point \(m_\pi = m_\sigma  = 0\), the \(Sp(N_f)\)-invariant field is the conformal Goldstone that has velocity \(c_{1} = 1/\sqrt{3}\).
Causality imposes a bound on the velocity of the other multiplet of Goldstones, \(c_{\smallday} \le 1\), but the precise value is not a priori fixed by the symmetries and depends on the details of the model.
To discuss them, we have to expand the \ac{eft} at quadratic order around the ground state, which we will do in the following section.

\section{Explicit second-order expansion}%
\label{sec:second-order}

In the previous section we have made a prediction for the low-energy spectrum over the ground state in Eq.~\eqref{eq:ground-state}.
Here we will verify it with an explicit computation of the dispersion relations at quadratic order in the fluctuations.

First we expand the field at quadratic order around the vacuum \(\Sigma_0\):
\begin{equation}
  \Sigma(t,x) = e^{i v(t,x)} \Sigma_0 e^{i v(t,x)^t},
\end{equation}
where
\begin{equation}
  v(t,x) = \twobytwo{\pi(t,x)}{0}{0}{-\pi(t,x)^t},
\end{equation}
and the fields \(\pi(t,x)\) satisfy \(\pi(t,x) \mathcal{I} - \mathcal{I} \pi(t,x)^t = 0\).
These are the generators of the broken global symmetries, encoding the coset \(U(N_f)/Sp(N_f)\).

Let us consider the terms in the action in Eq.~\eqref{eq:mu-Lagrangian} separately.
\begin{itemize}
\item The kinetic term gives
  \begin{equation}
    \ev*{\del_\mu \Sigma \del^\mu \Sigma} = 8 \sin[2](\varphi) \ev*{\del_\mu \pi \del^\mu \pi}.
  \end{equation}
  It is convenient to decompose \(\pi(t,x)\) on an orthonormal basis,
  \begin{equation}
    \pi(t,x) = \sum_{A = 0}^{\dim(\smallday)} \pi_A(t,x) T^A,
  \end{equation}
  where the generators \(T^A\) are normalized as
  \begin{equation}
    \ev*{T^A T^B} = \frac{1}{2} \delta^{AB} ,
  \end{equation}
  and in particular \(T^0 = 1/\sqrt{2 N_F} \Id_{N_f}\) (remember that we are only summing over the broken generators in \(u(N_f)\) that correspond to the coset \(U(N_f)/Sp(N_f)\)).
  Then we find
  \begin{equation}
    \ev*{\del_\mu \Sigma \del^\mu \Sigma} = 4 \sin[2](\varphi) \sum_{A=0}^{\dim(\smallday)} \del_\mu \pi_A \del^\mu \pi_A.
  \end{equation}
\item The term linear in \(\mu\) gives
  \begin{equation}
    \ev*{B \Sigma \del_0 \Sigma^\dagger} = -2 i \sin[2](\varphi) \ev*{\del_0 \pi},
  \end{equation}
  where we have used the fact that \(e^{iv}\) commutes with the baryonic charge.
  If we decompose \(\pi\) on our basis of the \(T^A\), we see that only the zeroth component survives:
  \begin{equation}
    \ev*{B \Sigma \del_0 \Sigma^\dagger} = -2 i \sin[2](\varphi) \ev*{\del_0 \pi} = - 2 i \sqrt{2N_f} \sin[2](\varphi) \del_0 \pi_0 .  
  \end{equation}
  This term is coupled to the fluctuations of the dilaton, which we expand as \(\sigma(t,x) = \sigma_0 + \hat \sigma(t,x)\).
\item The other terms do not depend on the derivatives of \(\Sigma\) and by definition do not contribute to the action for the pions.
  There is nonetheless a mass term for the fluctuations of the dilaton, which is a massive pseudo-Goldstone field, given by
  \begin{align}
    \ev*{B \Sigma_0 B \Sigma_0^\dagger} + \ev*{B^2} = N_f \sin[2](\varphi) , \\
    \ev*{M \Sigma_0} = 2 N_f \cos(\varphi) .
  \end{align}
\end{itemize}
The final result is that the action at quadratic order in the fluctuations is
\begin{equation}
 \frac{L[\pi, \hat \sigma]}{4 v^2 e^{-2 \sigma_0 f} \sin[2](\varphi)} =    \del_\mu \pi_0 \del^\mu \pi_0 - 2 \mu \sqrt{N_f} f \hat \sigma \del_0 \pi_0 - M_\sigma^2 \hat \sigma^2 + \sum_{a=1}^{\dim(\smallday)} \del_\mu \pi_a \del^\mu \pi_a ,
\end{equation}
where 
\begin{equation}
  M_\sigma^2 =  N_f f^2 \mu^2 + \frac{f^2 \mu^2 \left(-16 k_{4/3}^3 \mu^2 v^2 e^{2 f \sigma  y}+m_\pi^4 N_f (y^2-4) e^{6 f \sigma }+4 \mu^4 N_f e^{2 f \sigma  (y+1)}\right)}{2 \mu^4 e^{2 f \sigma  (y+1)}-2 m_\pi^4 e^{6 f \sigma }}.
\end{equation}
This is the effective mass of the dilaton, which is dominated by the fixed charge (via the parameter \(\mu\)) and remains large also for \(m_\sigma = 0\).
This is the behavior expected for a theory with an isolated fixed point~\cite{Orlando:2019skh}, as opposed to a supersymmetric system with a moduli space~\cite{Hellerman:2017sur,Hellerman:2018xpi}.

The massless fields \(\pi_a\), \(a = 1, \dots , \dim(\ydiagram{1,1})\) decouple from the rest of the system and have a linear dispersion relation with velocity \(c_{\smallday} = 1\).
The analysis of the symmetries had not been enough to predict this number.
The key observation is that the fields \(\pi_a\) do not enter the expansion of the linear term \(\ev*{B \Sigma \del_0 \Sigma^\dagger}\).
Geometrically speaking, this is the term that leads to the pre-symplectic form required to have type~II Goldstone fields~\cite{Watanabe:2012hr}.

The mode \(\pi_0\) is coupled to \(\hat \sigma\) and their inverse propagator reads
\begin{equation}
  \eval{D^{-1}}_{\pi_0, \hat \sigma} = \pmqty{ \omega^2 - k^2 & i \omega \mu f \sqrt{2 N_f } \\  -i \omega \mu f \sqrt{2 N_f } & -M_\sigma^2}.
\end{equation}
At the conformal point, where \(m_\pi = m_\sigma = 0\), we have \(M_\sigma^2 = N_f f^2 \mu^2\) and the propagator describes the expected conformal Goldstone mode, with dispersion relation \(\omega = k/\sqrt{3}\).
In the general case  we can still write the dispersion as an expansion in the inverse baryonic charge.
Once more it is convenient to consider the boundary values  \(\gamma^* = 0\) and  \(\gamma^* = 1\) separately:
\begin{align}
  \gamma^* &= 0 : & \omega &= \frac{1}{\sqrt{3}} \pqty{1 - \frac{N_f^{7/3}}{8 k_{4/3}^5} \frac{m_\pi^4}{v^2}  \pqty{\frac{ V}{Q}}^{2/3} + \dots } k ,\\
  \gamma^* &= 1 : & \omega &= \frac{1}{\sqrt{3}} \pqty{1 - \frac{N_f^{2/3}}{3 k_{4/3}} \pqty{ \frac{m_\sigma^2}{2f^2} + \frac{2 N_f^2}{ k_{4/3}^3} m_\pi^4  } \pqty{\frac{ V}{Q}}^{4/3} + \dots } k .
\end{align}

The leading quantum contribution to the energy is given by the Casimir energy of the Goldstones.
On a general manifold we can evaluate it using for example a zeta-function regularization:
\begin{equation}
  E_{\text{Casimir}} = \frac{1}{2} \pqty{c_1 + \frac{(N_f + 1 )(N_f -2)}{2} c_{\smallday}} \zeta(-1/2| \mathcal{M}^3) .
\end{equation}
The zeta function has a pole in \(s=-1/2\), which needs to be regulated, leading to a scheme-dependent result.
In fact, since the value of \(E_{\text{Casimir}}\) is  \(Q\)-independent, it will simply be absorbed by the Wilsonian parameter \(\tilde k_0\) in the energy of the ground state and the final result remains incalculable within the \ac{eft}.

As promised in Section~\ref{sec:EFT}, the quantum corrections only enter at \textsc{nnlo}.
Our main results in Eq.~(\ref{eq:dimension-small-gamma}) and Eq.~(\ref{eq:dimension-big-gamma}) are protected against corrections in \(1/Q\). Again, the semiclassical approach proves to be very effective in computing the operator dimensions close to the fixed point.

\section{Conclusions and outlook}%
\label{sec:conclusions}

We have analyzed the large-charge limit of  $Sp(N)$ gauge theories with $N_f$ Dirac fermions transforming in the fundamental representation of the gauge group near the lower boundary of the conformal window~\cite{Cacciapaglia:2020kgq}. The latter has been modeled by an \ac{eft} featuring the Goldstones of the theory augmented by a dilaton-like state. The large-charge approach   provides a controlled way to analyze strongly coupled near-conformal dynamics via a semi-classical expansion within the \ac{eft}. Having another tool at our disposal is invaluable given that unraveling the near-conformal dynamics has proven a formidable task, both numerically and analytically. 
 
After having introduced the fixed baryon charge of the theory and put the theory on a non-trivial background geometry  we have determined the spectrum and its symmetries,  disentangling the explicitly from the spontaneously broken symmetries.
Using the state-operator correspondence we have computed the corrections to the large-charge quasi-anomalous dimension \(\Delta\) as function of the dilaton and fermion mass as well as the background geometry:
\begin{itemize}
\item for small anomalous dimension of the fermion-mass operator  \( \gamma^* \ll 1\), we have
  \begin{multline}
    \frac{\Delta}{\Delta^*}  =
    1 - \frac{1}{N_f k_{4/3}^5} \pqty{\frac{9m_{\pi}^2}{64 \pi v}}^2  \bqty{1 - \gamma^* \log(\frac{3 \rho^{2/3}}{32 N_f  v^2  \pi^2 k_{4/3}} ) } \pqty{\frac{2 \pi^2}{\rho}}^{2/3}  \\
    - \bqty {\frac{16 \pi^2}{9} N_f k_{2/3} v^2  m_\sigma^2  %
    - \gamma^* \frac{4N_f^2}{3 k_{4/3}^6} \pqty{ \frac{9m_{\pi}^2}{64 \pi v}}^2 \pqty{ \frac{5N_f^2}{2k_{4/3}^4} \pqty{\frac{9m_{\pi}^2 }{64 \pi v}}^2  - k_{2/3}  }} \pqty{\frac{2 \pi^2}{\rho}}^{4/3} \log(Q) ;
  \end{multline}
\item for large anomalous dimensions \((1- \gamma^*) \ll 1\) we find
  \begin{equation}
    \label{eq:dimension-big-gamma}
    \frac{\Delta}{\Delta^*} = 1 - \bqty{\frac{16 \pi^2}{9} N_f k_{2/3} v^2  m_\sigma^2 + ( 1 - \gamma^*) \frac{9 m_\pi^4 }{64 k_{4/3}^4}   } \pqty{\frac{2 \pi^2}{\rho}}^{4/3} \log(Q) ,
  \end{equation}
  where \(\Delta^*\) is the conformal dimension at the fixed point and \(\rho\) the charge density.
\end{itemize}
The background geometry only enters via the charge density and  at very large charge the effect of the deformations from conformality is suppressed.
For \(m_\pi = 0\) we reproduce the results in~\cite{Orlando:2019skh}.
These results allow novel ways to explore near-conformal dynamics by offering, for example, independent tests for the existence of the dilaton and its impact. 

\medskip
We can envision a number of interesting further directions ranging from formal to phenomenological.  On the formal side one can extend the analysis to the \ac{qcd} ($SU(3)$) near-conformal window by replacing the baryon charge with the isospin charge~\cite{Splittorff:2000mm,Lenaghan:2001sd}. Generalizations to the conformal window of generic $SU(N)$ gauge theories with different matter representations are also highly relevant~\cite{Sannino:2004qp,Dietrich:2006cm}, one such example are \acl{gNJL} near-conformal theories~\cite{Rantaharju:2019nmh}. Another interesting avenue is the study of the effects of the topological sector of the underlying gauge theories encoded in the theta-angle~\cite{Witten:1979vv,Veneziano:1979ec,DiVecchia:1979pzw} of the theory whose effective description at the effective Lagrangian level~\cite{Rosenzweig:1979ay,DiVecchia:1980yfw,Crewther:1979rs,Witten:1980sp} (including a scalar isosinglet) was summarized in~\cite{DiVecchia:2013swa}.  
In order to further prepare for lattice investigations it would be also relevant to extend our study to take advantage of  finite volume computations~\cite{Gasser:1986vb,Gasser:1987ah,Leutwyler:1992yt} recently adapted to include the dilaton state in~\cite{Brown:2019ipr}. 
 
Additionally, our studies naturally fit into the radial quantization program, see~\cite{Brower:2018szu,Brower:2016vsl,Brower:2016moq,Brower:2015zea,Brower:2014daa} for recent attempts.  
 
At the same time the large and small charge limit investigated have relevant phenomenological applications~\cite{Badel:2019oxl} for bright and dark extensions of the standard model featuring composite dynamics. One example is investigating the physics of compact objects made by composite dark particles~\cite{Kouvaris:2019nzd,Chang:2018bgx,Eby:2015hsq,Hui:2016ltb} as also summarized in~\cite{Tulin:2017ara,Battaglieri:2017aum}. In principle one can use our \ac{eft} also to investigate \ac{qcd} in extreme matter conditions augmented by a sigma-like state~\cite{Baym:2017whm}.

\section*{Acknowledgments}

D.O. acknowledges partial support by the \textsc{nccr 51nf40--141869} ``The Mathematics of Physics'' (Swiss\textsc{map}).
The work of S.R. is supported by the Swiss National Science Foundation under grant number \textsc{pp00p2\_183718/1}.
The work of F.S. is partially supported by the Danish National Research Foundation under grant number \textsc{dnrf:90}. 

\newpage

\setstretch{1}

\printbibliography{}

\end{document}